\let\ifarxiv=\iffalse    
\pdfoutput=1
\documentclass[12pt,a4paper]{article} 

\ifarxiv\ifnum\pdfoutput=1\else
\PassOptionsToPackage{draft}{graphicx}
\fi\fi

\usepackage{amsmath,amssymb,bbm}
\usepackage{color}
\usepackage{graphicx}
\usepackage{multirow}
\usepackage{listings}	
\usepackage{slashed} 
\usepackage{multicol}
\usepackage{hhline}
\usepackage{array} 
\usepackage{pdflscape}  
\usepackage{nicefrac}		
\usepackage[makeroom,Smaller]{cancel}
\usepackage{ulem}
\usepackage{verbatim} 
\usepackage{tikz}
\usepackage{leftidx}
\usepackage{yfonts}
\DeclareMathAlphabet{\mathpzc}{OT1}{pzc}{m}{it}

\usepackage[a4paper,text={173mm,216mm},centering]{geometry}

\usepackage[font=small,labelfont=bf,width=0.85\textwidth]{caption}

\usepackage[breaklinks,bookmarks=true,hyperfigures=true]{hyperref}
\usepackage[nosort]{cite} %
\usepackage[bulletsep]{collref} 

\allowdisplaybreaks[3]

\makeatletter
\let\old@startsection=\@startsection
\renewcommand{\@startsection}[6]{\old@startsection{#1}{#2}{#3}{#4}{#5}{#6\mathversion{bold}}}
\makeatother

\makeatletter
\newlength{\apb@width}
\newcommand{\autoparbox}[2][c]{\settowidth{\apb@width}{#2}\parbox[#1]{\apb@width}{#2}}

\makeatother

\ifx\hypersetup\sadfkjashdfkxja\newcommand\hypersetup[1]{}\newcommand{\texorpdfstring}[2]{#1}\fi

\hypersetup{plainpages=false}
\hypersetup{pdfpagemode=UseNone}
\hypersetup{bookmarksnumbered=true}
\hypersetup{pdfstartview=FitH}
\hypersetup{colorlinks=false}
\hypersetup{citebordercolor={.5 1 .5}}
\hypersetup{urlbordercolor={.5 1 1}}
\hypersetup{linkbordercolor={1 .7 .7}}

\ifx\href\asklfhas\newcommand{\href}[2]{#2}\fi

\makeatletter
\def\Left#1#2\Right{\begingroup%
   \def\ts@r{\nulldelimiterspace=0pt \mathsurround=0pt}%
   \let\@hat=#1%
   \def\sht@im{#2}%
   \def\@t{{\mathchoice{\def\@fen{\displaystyle}\k@fel}%
          {\def\@fen{\textstyle}\k@fel}%
          {\def\@fen{\scriptstyle}\k@fel}%
          {\def\@fen{\scriptscriptstyle}\k@fel}}}%
   \def\g@rin{\ts@r\left\@hat\vphantom{\sht@im}\right.}%
   \def\k@fel{\setbox0=\hbox{$\@fen\g@rin$}\hbox{%
      $\@fen \kern.3875\wd0 \copy0 \kern-.3875\wd0%
      \llap{\copy0}\kern.3875\wd0$}}%
      \def\pt@h{\mathopen\@t}\pt@h\sht@im%
      \Right}%
\def\Right#1{\let\@hat=#1%
   \def\st@m{\mathclose\@t}%
   \st@m\endgroup}
\makeatother

\makeatletter
\def\mr@ignsp#1 {\ifx\:#1\@empty\else #1\expandafter\mr@ignsp\fi}%
\newcommand{\multiref}[1]{\begingroup
\xdef\mr@no@sparg{\expandafter\mr@ignsp#1 \: }%
\def\mr@comma{}%
\@for\mr@refs:=\mr@no@sparg\do{\mr@comma\def\mr@comma{,}\ref{\mr@refs}}%
\endgroup}
\makeatother

\renewcommand{\eqref}[1]{(\multiref{#1})} 

\newcommand{\hypref}[2]{\ifx\href\asklfhas #2\else\href{#1}{#2}\fi}

\newcommand{\appref}[1]{App.~\multiref{#1}}

\newcommand{\figref}[1]{Fig.~\multiref{#1}}

\newcommand{\vev}[1]{\langle#1\rangle}

\newcommand{\bra}[1]{\langle #1|}	

\newcommand{\ket}[1]{|#1\rangle}

\newcommand{\be}{\begin{equation}}
\newcommand{\ee}{\end{equation}}
\newcommand{\ba}{\begin{aligned}}
\newcommand{\ea}{\end{aligned}}
\newcommand{\bea}{\begin{eqnarray}}
\newcommand{\eea}{\end{eqnarray}}
\newcommand{\bean}{\begin{eqnarray*}}
\newcommand{\eean}{\end{eqnarray*}}

\renewcommand{\l} {\lambda}

\renewcommand{\lll}{\sqrt{\lambda}}

\renewcommand{\L}{\Lambda}

\renewcommand{\a} {\alpha}

\renewcommand{\b} {\beta}

\renewcommand{\d} {\delta}

\newcommand{\s} {\sigma}

\newcommand{\m}{\mu}
\newcommand{\n}{\nu}

\renewcommand{\k}{\kappa}
\newcommand{\om}{\omega}

\newcommand{\f}{\frac}

\renewcommand{\sf}[2]{{\textstyle\frac{#1}{#2}}}		

\newcommand{\calH}{\mathcal{H}}

\def\v{\vec}

\newcommand{\vXr}{\vec{X}_{\rm re}}
\newcommand{\vXi}{\vec{X}_{\rm im}}

\newcommand{\vXz}{\vec{X}_{0}}

\newcommand{\vx}{\vec{x}}
\newcommand{\vxpr}{\vec{x}^{\,}{}'{}}
\newcommand{\vxz}{\vec{x}_{0}}

\newcommand{\cx}{\mathpzc{x}}
\newcommand{\vcx}{\vec{\mathpzc{x}}}
\newcommand{\vcxpr}{\vec{\mathpzc{x}}^{\,}{}'{}}
\newcommand{\vcxz}{\vec{\mathpzc{x}}_{\,0}}

\newcommand{\vy}{\vec{y}}
\newcommand{\vypr}{\vec{y}^{\,}{}'{}}
\newcommand{\vyz}{\vec{y}_{0}}

\newcommand{\cy}{\mathpzc{y}}
\newcommand{\vcy}{\vec{\mathpzc{y}}}
\newcommand{\vcypr}{\vec{\mathpzc{y}}^{\,}{}'{}}
\newcommand{\vcyz}{\vec{\mathpzc{y}}_{0}}

\newcommand{\vPr}{\vec{P}_{\rm re}}
\newcommand{\vPi}{\vec{P}_{\rm im}}

\newcommand{\vPz}{\vec{P}_{0}}

\newcommand{\vpx}{\vec{p}_x}
\newcommand{\vpy}{\vec{p}_y}

\newcommand{\vpxz}{\vec{p}_{x,0}}
\newcommand{\vpyz}{\vec{p}_{y,0}}

\newcommand{\vcpx}{\vec{\mathpzc{p}}_\cx}
\newcommand{\vcpy}{\vec{\mathpzc{p}}_\cy}

\newcommand{\vcpxz}{\vec{\mathpzc{p}}_{\cx,0}}
\newcommand{\vcpyz}{\vec{\mathpzc{p}}_{\cy,0}}

\renewcommand{\Xi}{X_{\rm im}}

\newcommand{\alg}[1]{\mathfrak{#1}}

\newcommand{\dd}{{\rm d}}	
\newcommand{\Tr}{\mathop{\mathrm{Tr}}}		

\newcommand{\ord}[1]{\mathcal{O}(#1)}
\newcommand{\Ord}[1]{\mathcal{O}\left(#1\right)}

\newcommand{\p}{\partial}

\newcommand{\Reals}{\ensuremath{\mathbb{R }}}

\definecolor{pink}{rgb}{0.7,0,0.7}

\definecolor{grey}{rgb}{0.4,0.4,0.5}
\definecolor{darkgreen}{rgb}{0,0.5,0}
\definecolor{darkred}{rgb}{0.6,0.0,0}
\definecolor{lightbrown}{rgb}{1,0.9,0.8}
\definecolor{brown}{rgb}{0.6,0.3,0.3}
\definecolor{darkblue}{rgb}{0,0,0.8}
\definecolor{darkmagenta}{rgb}{0.5,0,0.5}

\newcommand{\calE}{{\mathcal{E}}}		

\def\dotuline{\bgroup
  \ifdim\ULdepth=\maxdimen  
   \settodepth\ULdepth{(j}\advance\ULdepth.4pt\fi
  \markoverwith{\begingroup
  \advance\ULdepth0.08ex
  \lower\ULdepth\hbox{\kern.15em .\kern.1em}%
  \endgroup}\ULon}

\def\dashuline{\bgroup
  \ifdim\ULdepth=\maxdimen  
   \settodepth\ULdepth{(j}\advance\ULdepth.4pt\fi
  \markoverwith{\kern.15em
  \vtop{\kern\ULdepth \hrule width .3em}%
  \kern.15em}\ULon}

\long\def\symbolfootnote[#1]#2{\begingroup
\def\thefootnote{\fnsymbol{footnote}}\footnote[#1]{#2}\endgroup}

\begin{document}



\begingroup\raggedleft\footnotesize\ttfamily
HU-EP-13/47\\
\endgroup
\begingroup\raggedleft\footnotesize\ttfamily
HU-Mathematik-2013-19\\
\endgroup
\begingroup\raggedleft\footnotesize\ttfamily
TCDMATH 13-12\\
\vspace{2mm}
\endgroup

\begin{center}
{\Large\bfseries
Static Gauge and Energy Spectrum of Single-mode Strings in $\bf AdS_{5}\times S^{5}$
\par}%
\vspace{5mm}

\begingroup\scshape\large
Sergey Frolov${\,}^{\star}$\footnote{ 
On leave from Trinity College Dublin. 
Correspondent fellow at
Steklov Mathematical Institute, Moscow.}, Martin Heinze${\,}^{\ast}$, \\
George Jorjadze${\,}^{\ast,\dag}$ and Jan Plefka${\,}^{\ast}$
\endgroup

\vspace{3mm}
\textit{${}^{\star}$ Institut f\"ur Mathematik und Institut f\"ur Physik, Humboldt-Universit\"at zu Berlin, \phantom{$^\S$}\\
IRIS Adlershof, Zum Gro{\ss}en Windkanal 6, 12489 Berlin, Germany}
\\ \vspace{2.5mm}
\textit{${}^{\star}$ Hamilton Mathematics Institute and School of Mathematics, 
\phantom{$^\S$}\\
Trinity College, Dublin 2, Ireland}\\ \vspace{2.5mm}
\textit{${}^{\ast}$ Institut f\"ur Physik, Humboldt-Universit\"at zu Berlin, \phantom{$^\S$}\\
Newtonstra{\ss}e 15, D-12489 Berlin, Germany}\\ 
\vspace{2.5mm}
\textit{${}^{\dag}$ Razmadze Mathematical Institute of JSU and FU of Tbilisi,\\
University Campus, David Aghmashenebeli Alley, Tbilisi, Georgia} \\[0.1cm]
\texttt{\small frolovs@maths.tcd.ie,\\ \{martin.heinze, george.jorjadze, jan.plefka\}@physik.hu-berlin.de\phantom{\ldots}}
\vspace{8mm}


\textbf{Abstract}\vspace{5mm}\par
\begin{minipage}{14.7cm}
Motivated by the notorious difficulties in determining the first quantum corrections to the
spectrum of short strings in $AdS_5 \times S^5$ from first principles, 
we study closed bosonic strings in this background employing a static gauge.
In this gauge the world-sheet Hamiltonian density is constant along the extension of the string
and directly proportional to the square of the spacetime energy.
We quantize this system in  a minisuperspace approach, in which we consider only a single
$AdS_5$ string mode excitation next to an arbitrary particle like zero-mode contribution
in the full $AdS_5 \times S^5$ background.   
We determine the quantum spectrum using this method to the next-to-next-to-leading
 order in the large 't Hooft
coupling expansion. We argue for an ordering prescription which should arise from supersymmetrization and indeed recover the integrability
based predictions for the spectrum of the lightest
excitation, dual to the Konishi field scaling dimensions. The higher excitations fail
to agree, but this is shown to be a consequence of the string mode truncation employed.
Despite this simple setup, our system reveals intriguing features, such as a close connection to particles in $AdS_6$, classical integrability and preservation of 
the isometries of $AdS_5 \times S^5$ at the quantum level. 

\end{minipage}\par
\end{center}
\newpage

\section{Introduction and Conclusions} \label{sec:Intro}

One of the classic problems in the maximally supersymmetric
AdS/CFT correspondence is the computation of the quantum superstring excitation
spectrum in a suitable form of perturbation theory.
The dual of short strings moving in $AdS_{5}\times S^{5}$  spacetime are
the finite length non-protected local operators of
4d $\mathcal{N}=4$ super Yang-Mills theory
with the length-two Konishi operator $\Tr(\Phi^{I}\Phi^{I})$ being the first representative.
In fact, for the Konishi operator the anomalous scaling dimension 
 has been computed for small 't Hooft coupling $\lambda\ll 1$ in
field  perturbation theory up to an impressive 5-loop order \cite{Anselmi:1996dd,
 Bianchi:2000hn, Eden:2000mv, Eden:2000vb, Arutyunov:2000im, Fiamberti:2007rj,Fiamberti:2008sh, Velizhanin:2008jd, Eden:2012fe}. The motivation for going to such high orders arose 
 from the enormous progress in understanding the hidden integrable system
behind the spectral problem of this AdS/CFT duality pair (for a recent review see \cite{Beisert:2010jr}).
Here, the computation of the Konishi scaling dimension has become something of a
testing ground for the application  of integrability techniques going beyond the asymptotic Bethe ansatz \cite{Beisert:2006ez} in the form of the thermodynamic Bethe ansatz for the mirror model 
\cite{Arutyunov:2009zu,Arutyunov:2009ur,Bombardelli:2009xz,Gromov:2009bc}
or the Y-system \cite{Gromov:2009tv}. The assumption of integrability is
powerful enough to evaluate the Konishi scaling dimension to even higher orders \cite{Bajnok:2009vm,Arutyunov:2010gb,
Balog:2010xa, Leurent:2012ab, Bajnok:2012bz}, with the present 
record being set at eight \cite{Leurent:2013mr, Gromov:2013pga} or even nine loops \cite{VolinTalk}. 
The integrability based results in
principle can yield the scaling dimensions of short operators dual to the spectrum of short excited $AdS_{5}\times S^{5}$ strings for {\it any} value of the coupling $\lambda$ and in
particular also at strong coupling, $\lambda\gg 1$, i.e.~the stringy regime. 
In this limit, thermodynamic
Bethe ansatz methods have been applied to find numerically the spectrum of the first string
excitations, and to extract the first few coefficients in the strong coupling expansion up to the next-to-next-to-leading order 
\cite{Gromov:2009zb, Frolov:2010wt, Gromov:2011bz,Frolov:2012zv}.
Of course, also these results hinge on the assumption of integrability in the AdS/CFT system.

Therefore also at strong coupling it is desirable to scrutinize the integrability
assumption through an independent first principle computation of the $AdS_{5}\times S^{5}$ string
spectrum at least for the lowest excitations. This amounts to the perturbative
study of the quantum superstring in the large
effective string-tension ($T_0 R^2= \sqrt{\lambda}/ 2 \pi$) expansion.
 Here however, string theory lags behind
the gauge theory as this perturbative
approach has not been developed in a completely
satisfactory fashion to date. The single string-mode
spectrum is known from first principles at leading order in a $1/\sqrt{\lambda}$ expansion
\cite{Gubser:1998bc}, where it is given by the flat-space
result, and the next-to-leading order term, related to the excitations of the string
zero-modes \cite{Passerini:2010xc}, capturing the leading center-of-mass or particle like
dynamics. The first interesting term sensitive to stringy effects
in the curved $AdS_{5}\times S^{5}$ background, however, lies at the next-to-next-to leading order 
\be
E_{(Q,m)}= c_{-1}\sqrt[4]{m^{2}n^2\lambda} + c_{0} + \frac{c_{1}}{\sqrt[4]{m^{2}n^2\lambda}} +
\frac{c_{2}}{\sqrt{m^{2}n^2\lambda}}  + \frac{c_{3}}{(m^{2}n^2\lambda)^{3/4}} + \ldots~ , \label{eq:In_EnExpand}
\ee
Generally, the coefficients are functions of the string-mode number $m$, 
as well as the number of excitations $n$ of the $m$th mode
and  $Q$ denotes the conserved global charges of the state, which for example could be the $S^{5}$ total angular momentum $J$ or the total $AdS_{5}$ spin $S$ of the state.

The above mentioned direct
near-flat space string and zero-mode computations yield 
\be
c_{-1}=2~, \qquad c_{0}\in\mathbb{Z}/2 \quad \text{depending on the state.}
\ee
In fact
the integrability based numerical results of \cite{Frolov:2012zv} confirm this and
moreover yield 
\be 
c_{0}= 0\, , \qquad
c_{1}=\frac{J^{2}}{4} + \frac{m(3-m)}{2}~, \qquad m=1,2,3,4~ , \label{eq:In_c2Frolov}
\ee
for gauge theory operators from the $\alg{sl}(2)$ sector with angular momentum $J$ which
should have $n=1$ in the string language.
For the lightest state in this series
being a member of the Konishi multiplet one has $J=2$ and $m=1$
yielding $c_{1}=2$.
It is still an open challenge for a direct string computation to confirm this
prediction.\footnote{In \cite{Vallilo:2011fj}, the term $c_{1}$ has been claimed to be computed from first principles using pure spinor formalism.}

On the other hand the situation for a string based computation employing the
``semiclassical'' approach suggested in \cite{Roiban:2009aa,Tirziu:2008fk}
and further applied in 
\cite{Gromov:2011de,Beccaria:2011uz,Roiban:2011fe,Gromov:2011bz,Beccaria:2012xm,Beccaria:2012kp} is much better developed. Here one identifies classical bosonic string solutions carrying
definite charges $Q$.
In a semi-classical fluctuation quantization about such backgrounds the spacetime
energy can be reliably computed in an $1/\sqrt{\lambda}$ string-worldsheet loop computation
for $Q\propto \sqrt{\l}$. The obtained results are then continued to the relevant
small charges for the short string spectral problem, $Q \propto 1$, assuming no order of limits
ambiguities. Using this the above quoted value for the Konishi-multiplet $c_{1}=2$ has
been reproduced. Alternative methods for establishing the strong coupling spectrum based
on the algebraic curve approach, exploiting the classical integrability of the world-sheet
theory to extrapolate to the one-loop order, have also been able to reproduce these results
\cite{Gromov:2011de,Basso:2011rs,Gromov:2011bz,Beccaria:2012kp}. 

In this work we report on an alternative method to find the quantum excitation
spectrum for the lowest $AdS_{5}\times S^{5}$ string excitation in a direct
approach employing a static gauge paired with a minisuperspace approximation. 
 The distinct advantage of a static gauge $X^{0}\sim E\tau$
for the AdS-string is that the worldsheet Hamiltonian density is polynomial in
the fields,  independent of the worldsheet space coordinate $\sigma$
and in fact proportional to the target-space energy squared operator $E^{2}$.
It thus has a simpler structure than the more commonly used uniform light-cone gauge Hamiltonian
\cite{Arutyunov:2005hd,Frolov:2006cc,Arutyunov:2009ga} being of square-root form (like
the Nambu-Goto action) and directly proportional to $E-J$. The drawback compared to the
light-cone gauge is that one cannot solve all constraints at the classical level and
is left with the quantization of a constrained system.
 Recently the static gauge quantization of the bosonic string in flat spacetime
was achieved \cite{Jorjadze:2012iy} and in this work we build upon these results.
We introduce the static gauge for the $AdS_5\times S^5$ bosonic string and work out the  Hamiltonian and all conserved charges in the classical
theory. In order to find the perturbative quantum spectrum for large ${\lambda}$
we consider a novel approach. We consider a solution which carries
a single string-mode  excitation on $AdS_{5}$ and arbitrary zero-mode excitations.
The dynamics is that of a massive particle in AdS-space with a mass
determined by the internal single-mode
string oscillations as well as the $S^5$ total angular momentum $J$. It describes
a pulsating string whose center of mass oscillates within the confining AdS-space potential
and limits to previously known pulsating string solutions with vanishing zero mode
dependence. The solution is still rather general and parameterized by an 8+2 dimensional
phase space. It exists only in
$AdS_{n\geq 5}$ spaces, as the zero-modes and non zero-modes oscillate
in orthogonal spatial planes.
We then quantize this reduced dynamical system in what could be called a minisuperspace approach.
We also show that our quantization respects the bosonic isometry subgroup 
$SO(2,4)\times SO(6)$, which is non-manifest in the $SO(2,4)$ part due to the static gauge 
choice.
Although the computation is plagued by ordering ambiguities we argue for a natural
prescription which we expect to follow from supersymmetrizing the problem and implementing
the superconformal symmetry algebra at the quantum level. Our analysis leads to
a spectrum of the form
\be\label{eq:In_EnSpectrum}
E_{(N,m,n,J)}=2\,\l^{1/4}\sqrt{m\,n} + N-2 + \frac{10\,n^2-6\,n+4+{\cal M}^2_S}{4\,\l^{1/4}\sqrt{m\,n}}+\ord{\l^{-1/2}} \, ,
\ee
where $N$ accounts for the particle-like excitations in $AdS_{5}$ space, and ${\cal M}^2_S$ is the contribution due to the motion on the $S^5$ which for the bosonic string is equal to the SO(6) Casimir ${\cal M}^2_S = J(J+4)$. It is expected that the fermion contribution would modify the $J$ dependence to  ${\cal M}^2_S = J^2 +\ord{\l^{-1/4}}$.  
For the lightest stringy state with $N=0$, 
$m = n = 1$ and $J=0$ which is dual to the Konishi operator $\Tr \Phi^I\Phi^I$, we get
$c_{1}=2$ consistent with the integrability findings \cite{Gromov:2009zb,Frolov:2010wt, Frolov:2012zv}.

Lowest excitations of higher modes, $n=1$ but $m>1$, fail to agree with the integrability based results \eqref{eq:In_c2Frolov}.
This is not too surprising as the level-truncation we have effectively applied will not agree with
the full quantum fluctuation result as higher intermediate levels will become important.
We should stress once more that the above result pertains to a particular normal-ordering
prescription, which we consider to be natural. However, other ordering prescriptions
will change the numerical value of the next-to-next-to leading order term $c_{1}$ in (\ref{eq:In_EnSpectrum}).

This situation thus strongly asks for an inclusion of the fermionic degrees of freedom
for the Green-Schwarz superstring in $AdS_{5}\times S^{5}$ \cite{Metsaev:1998it,Arutyunov:2009ga}. 
For this the framework of static gauge, which then must be accompanied by a compatible 
$\kappa$-symmetry gauge
fixing condition, has still to be developed. 
 Finally, the question arises, whether the presented approach can be applied to other semiclassical string solutions. These challenging problems are left for future work.


\section{Closed bosonic string dynamics in static gauge} \label{sec:BSSG}

First we consider the static gauge approach to string dynamics in a generic static spacetime.
We use only a part of the constraints to exclude the time components of the canonical variables.
This relates the energy square functional to the Hamiltonian of the system and excludes negative norm states on the quantum level.
The remaining constraints specify the physical Hilbert space.
We then focus on $AdS_{N+1}\,\times\,S^M$, where we explore the dynamical integrals of the isometry group
and analyze their behavior at large coupling.

\subsection{String in a static spacetime} \label{subsec:SSST} 

Bosonic string dynamics in a curved spacetime with coordinates $X^\m$, $\m=0,1,...,D-1$, and
metric tensor $G_{\m\n}(x)$ is described by the Polyakov action
\be\label{eq:SSST_Polyakov}
S=-\frac{T_0}{2}\int  \mbox{d}\tau\,\mbox{d}\s\,\,
\sqrt{-h}\,h^{\a\b}\,\p_\a X^\m\,\p_\b  X^\n\,G_{\m\n}(x)~.
\ee
The worldsheet coordinates $(\tau,\s)$ are dimensionless and the string tension $T_0$ has the dimension of
inverse length square. In first order formalism, for a closed string, this action  is equivalent to
\be\label{eq:SSST_Action}
S=\int \mbox{d}\tau \int_0^{2\pi} \frac{\mbox{d}\s}{2\pi}\, \Big(P_\m\,\dot X^\m+
\frac{\xi_1}{2T}\left(G^{\m\n}\,
P_\m P_\n + T^2 G_{\m\n}\,X'{\,^\m} X'{\,^\n}\right)
+\xi_2\,\,P_\m X'^\m\Big)~.
\ee
Here, $P_\m$ are the momentum variables conjugate to $X^\m$, $T\equiv2\pi\,T_0$ and $\xi_1=\frac{1}{\sqrt{-h}\,h^{00}},$
$\xi_2=\frac{h^{01}}{h^{00}}$
play the role of Lagrange multipliers. Their variations provide the Virasoro constraints
\be\label{eq:SSST_Constr0}
G^{\m\n}\,P_\m P_\n + T^2 G_{\m\n}\,X'{\,^\m} X'{\,^\n}=0~,\quad\quad P_\m X'{\,^\m}=0~.
\ee

Let us consider a static spacetime. Its metric tensor can be brought into the form
\be\label{eq:SSST_GenMetric}
G_{\m\n}=\left(\begin{array}{cc}
              -\L & 0 \\
              0 & G_{kl}
            \end{array}\right)~,
\ee
with $\L>0$ and positive definite $G_{kl}$, $k, l=1,2,...,D-1$, both $X^0$-independent.
By Noether's theorem one then has a gauge invariant conserved energy
\be\label{eq:SSST_Energy}
E=- \int_0^{2\pi} \frac{\mbox{d}\s}{2\pi}\, ~P_0~.
\ee

The static or temporal gauge is defined by the gauge fixing conditions
\be\label{eq:SSST_GaugeFix}
X^0=-\a\,P_0\,\tau~, \qquad P_0^{\,\prime}(\tau,\s)=0~,
\ee
with a positive constant $\alpha$.
As the spacetime coordinates have the dimension of length
we take $\a=1/T$. In static spacetimes with an intrinsic length scale $R$, such as $AdS_{N+1}$,
one may rescale to dimensionless coordinates leading to an effective
$\alpha=1/(R^2 T)$.


In the static gauge the first order action \eqref{eq:SSST_Action}
reduces to\footnote{We neglect the boundary term related to the time derivative
$-\p_\tau\left(\frac{\a}{2}\,P_0^2\tau\right)$.}
\be\label{eq:SSST_RedAction}
S=\int \mbox{d}\tau \int_0^{2\pi}\frac{\mbox{d}\s}{2\pi}\,\left(P_k\,\dot X^k-
\frac{\a P_0^2}{2}\right)~,
\ee
where the square of the energy density is fixed by
the first Virasoro constraint in \eqref{eq:SSST_Constr0} and reads
\be\label{eq:SSST_EnSquared}
P_0^2={\L}\Big( G^{kl}P_k P_l+
T^2\,G_{kl}\,X'{\,^k} X'{\,^l}\Big)~.
\ee
Taking into account the second constraint in \eqref{eq:SSST_Constr0} and
the gauge fixing conditions \eqref{eq:SSST_GaugeFix},
we find that the Hamiltonian system \eqref{eq:SSST_RedAction}-\eqref{eq:SSST_EnSquared} has to be further reduced
to the constraint surface
\be\label{eq:SSST_Constr1}
{\cal H}'(\s)=0~,
\quad\quad {\cal{V}}(\s)\equiv P_k(\s) X'{\,^k}(\s)=0~,
\ee
where ${\cal H}(\s)=\frac{\a P_0^2}{2}$ is the Hamiltonian density in \eqref{eq:SSST_RedAction} and  we call ${\cal{V}}(\s)$
the level matching density. Thus, ${\cal H}(\s)$ has vanishing non zero-modes
 and on the constraint surface one gets with $H=\int^{2\pi}_0 \frac{d\sigma}{2\pi}\, \mathcal{H}(\sigma)$
\be\label{eq:SSST_EnSq&Ham}
E^2=\f{2H}{\a}=\int_0^{2\pi} \frac{\mbox{d}\s}{2\pi}\,\,{\L}\Big( G^{kl}P_k P_l+
T^2\,G_{kl}\,X'{\,^k} X'{\,^l}\Big)~.
\ee
In fact on the constraint surface the integrand in the above is constant.

The construction of independent canonical variables on the constraint surface \eqref{eq:SSST_Constr1}
is not an easy task even for Minkowski space with $G_{kl}=\d_{kl}$ and $\L=1$.
In this case the system \eqref{eq:SSST_RedAction}-\eqref{eq:SSST_EnSquared}
describes $(D-1)$ massless free fields in two dimensions and the constraints \eqref{eq:SSST_Constr1}
correspond to $L_n=0=\bar L_n$ for $n\neq 0$, where $L_n $ and $\bar L_n$ are the standard Virasoro generators of 2d free CFT.
It is therefore easier to first quantize the free-field theory and then
take into account the constraints \eqref{eq:SSST_Constr1} on the quantum level  by the equations
for the physical states $L_n|\psi_{{ph}}\rangle=0=\bar L_n|\psi_{{ph}}\rangle$  ($n>0$) \cite{Jorjadze:2012iy}.
To follow the same scheme for a generic static spacetime one has to analyze
the Poisson bracket structure of the constraints \eqref{eq:SSST_Constr1}.

The canonical relations $\{P_k(\s), X{^l}(\tilde\s)\}=2\pi\d_k^l\,\d(\s-\tilde\s)$
provide the Poisson brackets
\bea\nonumber
&&\{{\cal V}(\s) ,{\cal V}(\tilde\s)\}=-2\pi\left[{\cal V}\,'(\s)\,\delta(\s-\tilde\s)+
2{\cal V}(\s)\,\delta'(\s-\tilde\s)\right]~,
\\ [.4 em] \label{eq:SSST_ConstrAlgebra}
&&\{{\cal V}(\s),{\cal H}(\tilde\s)\}=-2\pi\left[{\cal H}'(\s)\,\delta(\s-\tilde\s)+
2{\cal H}(\s)\,\delta'(\s-\tilde\s)\right]~,
\\ [.4 em] \nonumber
&&\{{\cal H}(\s), {\cal H}(\tilde\s)\}=-2\pi \a^{2}[(\L^2(\s){\cal V}(\s))'\,\delta(\s-\tilde\s)+
2\L^2(\s){\cal V}(\s)\,\delta'(\s-\tilde\s)]~.
\eea
Equation \eqref{eq:SSST_Constr1}, thereby, defines the second class constraints
in Dirac's classification, as in the flat case, and they are preserved in dynamics
\be\label{eq:SSST_ConstrDynamics}
\{H,{\cal V}(\s)\}={\cal H}'(\s)~,\qquad \quad \{H,{\cal H}'(\s)\}=\a^{2}(\L^2(\s){\cal V}(\s))''~.
\ee

To quantize this system one has to realize the quantum version of these Poisson bracket relations
and use the constraint operators to select the physical Hilbert space as in the flat case.
The spectrum of the Hamiltonian on the physical states then will define the energy spectrum of the system
on the basis of the relation \eqref{eq:SSST_EnSq&Ham}.

\subsection{String in $AdS_{5}\times S^{5}$} \label{subsec:AdSS} 

Let us consider an $AdS_{N+1}\times S^M$ space with common radius $R$, where $AdS_{N+1}$
is realized as a $(N+1)$-dimensional hyperbola ${\cal Z}_A {\cal Z}^A=-R^2$ in $\Reals^{2,N}$ with embedding coordinates ${\cal Z}^A,$ $A = 0',0,1,\ldots,N$, and $S^M$ as $M$-dimensional sphere ${\cal Y}_I {\cal Y}^I = R^2$ in $\Reals^{M+1}$, with $I=1,\ldots,M+1$.

We parameterize the embedding coordinates of $\Reals^{2,\,N}$ by dimensionless $X^0,$ $X^a$ as
\be\label{eq:AdSS_AdSCoords}
{\cal Z}^{0'}=R \sqrt{1+\v{X}^2}\,\,\sin(X^0)~,\qquad {\cal Z}^{0}=R\sqrt{1+\v{X}^2}\,\,\cos(X^0)~,
\qquad {\cal Z}^a =R\,X^a~,
\ee
where $X^0$ is the $AdS_{N+1}$ time coordinate and $ \v{X}^2 \equiv X^b X^b$, $a, b =1,\ldots,N$.

For the spherical part one can use the coordinates of the stereographic projection
\be\label{eq:AdSS_SCoords}
{\cal Y}^{i}= \frac{R\,Y^i}{1+\v{Y}^2/4}~,\qquad {\cal Y}^{M+1} = R\,\frac{1-\v{Y}^2/4}{1+\v{Y}^2/4}~,
\ee
with $Y^i=X^{N+i}$ and $\v{Y}^2 \equiv Y^j Y^j$, $i, j = 1,\ldots,M$.

The induced metric on $AdS_{N+1}\times S^M$ takes the following block structure
\be\ba\label{eq:AdSS_IndMetric}
g_{\m\n}=\left(\begin{array}{ccc}
              -\L & 0 & 0\\
              0 & G_{ab}&0\\
              0 &0 &G_{ij}
            \end{array}\right)~,\qquad  \mbox{where}~~~~~~~~~~~~~~~~~~~~~~~~~~~ \\
\L=R^2(1+ \v{X}^2)~,\qquad  G_{ab}= R^2\left(\d_{ab}-\frac{X^a\,X^b}{1+\v{X}^2}\right)~,\qquad
G_{ij}= \frac{R^2\, \d_{ij}}{(1+\v{Y}^2/4)^2}~,\ \ \qquad
\ea\ee	
and the corresponding inverse matrices read
\be\label{eq:AdSS_InvMetric}
G^{ab}= \frac{1}{R^2}\left(\d_{ab}+ {X^a\,X^b}\right)~,\qquad
G^{ij}=\frac{1}{R^2} \left(1+\v{Y}^2/4\right)^2\,\d_{ij}~.
\ee
	
We denote the momentum variables conjugated to $X^a$ and $Y^i$  by $P_{a}$ and $ P_{Y^i}= P_{N+i}$, respectively. Taking into account the structure of the metric tensors in \eqref{eq:AdSS_IndMetric}, it is
convenient to treat $X^a$ and $P_a$ as vectors of $\Reals^N$, $Y^i$ and $P_{Y^i}$ as vectors of $\Reals^M$, and use
standard notations of vector algebra for $O(N)$ and $O(M)$ scalars: 
$\v{P}_X^{\,2}\equiv \v{P}^{\,2}\equiv P_aP_a$, $\v{P}_Y^2\equiv P_{Y^i}P_{Y^i}$, {\it etc.}~as in \eqref{eq:AdSS_AdSCoords} and \eqref{eq:AdSS_SCoords}.

Applying the static gauge \eqref{eq:SSST_GaugeFix}, with $\a=1/(R^2T)$, to the $AdS \times S$ string,
by \eqref{eq:SSST_EnSq&Ham} we obtain
\be\label{eq:AdSS_EnSq&Ham}
E^2={2\sqrt\l\, H}=\int_0^{2\pi}\frac{\mbox{d}\s}{2\pi}\,\,(1+ \v{X}^2)\left[ \v{P}^{\,2} + (\v{P}\cdot \v{X})^2 +
\l\left( \v{X}'^{\,2}-\frac{( \v{X} \cdot  \v{X}')^2}{1+ \v{X}^2}\right) +
{\cal M}_S^2\right]~,
\ee
where $\l\equiv R^4\,T^2$ is the 't Hooft coupling of the dual $\mathcal{N}=4$ super Yang-Mills
gauge theory
and ${\cal M}_S^2$ denotes the spherical part
\be\label{eq:AdSS_SHam}
{\cal M}_S^2=(1+\v{Y}^2/4)^2\,\v{P}_Y^{\,2}+\l\,\frac{\v{Y}'^{\,2}}{(1+\v{Y}^2/4)^2}~.
\ee

Analyzing the Hamiltonian system defined by \eqref{eq:AdSS_EnSq&Ham}-\eqref{eq:AdSS_SHam} one has to recall
that the integrand in \eqref{eq:AdSS_EnSq&Ham} is $\s$-independent and, in addition,
that the level matching density vanishes
\be\label{eq:AdSS_LevelMatch}
{\cal V}=\v{P}\cdot\v{X}'+\v{P}_{Y}\cdot\v{Y}'=0~.
\ee

The dynamical integrals related to the rotations in $\Reals^{2,N}$ and $\Reals^{M+1}$,
\be\label{eq:AdSS_IoM}
J_{A\,B}=\int_0^{2\pi}{\frac{\mbox{d}\s}{2\pi}\,}\,{ V}_{AB}^\mu P_\mu~, \qquad
 L_{I\,J}=\int_0^{2\pi}{\frac{\mbox{d}\s}{2\pi}\,}\,{ V}_{IJ}^\mu P_\mu~,
\ee
generate the isometry group SO$(2,N)\times$SO$(M+1)$ and they are dimensionless, as is the energy.
The index $\mu$ in \eqref{eq:AdSS_IoM} incorporates the three blocks $\m=(0,\,a,\,N+i)$,
$P_0=-E$ and ${V}_{AB}^\m$,
${ V}_{IJ}^\mu$ are the components of the Killing vector fields in $AdS_{N+1}\times S^M$,
\bea\label{eq:AdSS_VectFields}
{ V}_{AB}^0 = G^{00}({\cal Z}_B \p_0 {\cal Z}_A - {\cal Z}_A\p_0 {\cal Z}_B)~, \qquad
{ V}_{AB}^a = G^{ab}({\cal Z}_B\p_b {\cal Z}_A - {\cal Z}_A \p_b {\cal Z}_B)~,\\ [.4 em] \nonumber
{ V}_{IJ}^{i+N}=G^{ij}({\cal Y}_J\p_j {\cal Y}_I -{\cal Y}_I\p_j {\cal Y}_J)~.~~~~~~~~~~~~~~~~~~~~~~~~
\eea
Using then \eqref{eq:AdSS_AdSCoords}-\eqref{eq:AdSS_InvMetric} and the notation  $\om \equiv E/\sqrt\l$, we find
\be\label{eq:AdSS_SO24Rots}
{J}_{0'\,0} = E ~, \qquad \quad \qquad
 \qquad\ \ {J}_{a\,b}=\int_0^{2\pi}{\frac{\mbox{d}\s}{2\pi}\,}
\left( P_{a}\,X^b - P_{b}\,X^a\right)~,
\ee
\be\ba\label{eq:AdSS_SO24Boosts}
&&{J}_{a\,0'} =E\left(\int_0^{2\pi}\frac{\mbox{d}\s}{2\pi}\,\frac{X^a}{\sqrt{1+ \v{X}^2}}\right)
\cos\left(\om\,\tau\right)
-\left(\int_0^{2\pi}\frac{\mbox{d}\s}{2\pi}\,
\sqrt{1+\v{X}^2}\,\,P_{a}\right)\sin\left(\om\,\tau\right), ~
\\
&&{J}_{a 0} =-E\left(\int_0^{2\pi}\frac{\mbox{d}\s}{2\pi}\,\frac{X^a}{\sqrt{1+ \v{X}^2}}\right) \sin\left(\om\,\tau\right)
-\left(\int_0^{2\pi}\frac{\mbox{d}\s}{2\pi}\,
\sqrt{1+\v{X}^2}\,\,P_{a}\right)\cos\left(\om\,\tau\right),
\ea\ee
\be\label{eq:AdSS_SO6Rots}		
L_{i\,j} = \int_0^{2\pi}\frac{\mbox{d}\s}{2\pi} \left( P_{Y^i}\,Y^j - P_{Y^j}\,Y^i\right),
 \quad L_{i\,M+1} = \int_0^{2\pi}\frac{\mbox{d}\s}{2\pi} \Big(\big(1-\v{Y}^2/4\big)P_{Y^i}+\frac{1}{2}\,(\v{P}_Y \cdot\v{Y})Y^i\Big).
\ee

Our aim is to analyze the $AdS_5\times S^5$ string dynamics at large coupling $\l \gg 1$. The Hamiltonian
of the system defined by \eqref{eq:AdSS_EnSq&Ham} allows the expansion in powers of $1/\sqrt\l$
\be\label{eq:AdSS_HamExpand}
H=H^{(0)}+\frac{1}{\sqrt\l}\,H^{(1)}+\f{1}{\l}\,H^{(2)}+\dots ~,
\ee
which is easily obtained if one uses the rescaled phase space coordinates
\be\label{eq:AdSS_Rescale}
x^k=\l^{1/4}\,X^k~, \qquad \quad p_k=\l^{-1/4} P_k~,
\ee
where $X^k=(X^a,\,Y^i)$, $P_k=(P_a,\,P_{Y^i})$, $k=1,2,\dots ,9$. 
The leading term in \eqref{eq:AdSS_HamExpand} coincides with the static gauge string Hamiltonian in $10$-dimensional Minkowski space
\be\label{eq:AdSS_HamLead}
H^{(0)}=\frac{1}{2}\int_0^{2\pi}\frac{\mbox{d}\s}{2\pi}\left(p^2(\s)+x'^{\,2}(\s)\right)~.
\ee
The rotation generators $J_{a\,b}$ and $J_{i\,j}$ in \eqref{eq:AdSS_SO24Rots}-\eqref{eq:AdSS_SO6Rots} are invariant under the rescaling \eqref{eq:AdSS_Rescale}
and, therefore, they are $\l$-independent.
The generators  $J_{a\,0'}$ are expanded in powers of $1/\sqrt\l$ like the Hamiltonian. The leading terms of their expansion correspond
to four boosts in $x^a$ ($a=1,\dots , 4$) directions of the 10-dimensional Minkowski space.
Other symmetry generators in \eqref{eq:AdSS_SO24Rots}-\eqref{eq:AdSS_SO6Rots} are singular at $\l\rightarrow \infty$, however, after
their rescaling by the factor $\l^{-1/4}$, they also become analytic in $1/\sqrt{\l}$. It is easy to check that the
corresponding leading terms define the translation generators of 10-dimensional Minkowski space.
In fact, the zero-modes will have to be scaled differently than stated in (\ref{eq:AdSS_Rescale})
as was already noted in \cite{Passerini:2010xc} and will be discussed in detail in section 4.2.

With a supersymmetric extension, the leading order part is quantized without anomalies. This part of the symmetry generators and the commutation relations of the isometry group can be used as a basis for perturbative quantum calculations.


\section{Single-mode strings in $AdS_5 \times S^5$} \label{sec:OMSiAdSxS}

In this section we introduce a class of $AdS_5\times S^5$ string configurations with excited zero-modes
both in the AdS and the spherical parts and only one excited non zero-mode in the AdS part.
We call these configurations single-mode strings.
In the first part of the section we describe the Hamiltonian reduction to the single-mode ansatz,
where we find canonical physical variables of the system and verify the SO(2,4)$\times$SO(6) symmetry of single-mode strings.
Using a canonical transformation to new variables, which separates the non zero-mode part, we then
show the integrability of the single-mode configurations. They turn out to describe pulsating strings.

\subsection{Hamiltonian reduction to the single-mode ansatz} \label{subsec:HRed}

The Hamiltonian density of the $AdS_5\times S^5$ string is expressed in terms of O(4) and O(5) scalars
denoted by $(\v P$, $\v X)$ and $(\v{P}_Y$, $\v Y)$, respectively (see \eqref{eq:AdSS_EnSq&Ham}-\eqref{eq:AdSS_SHam}).
We introduce the above mentioned single-mode ansatz in the following form
\bea\label{eq:HRed_OMAnsatz}
&\v{X} (\tau,\s) = \v{X}_0(\tau) + \v{X}_+(\tau)\, e^{i m \s} + \v{X}_-(\tau)\, e^{-i m \s} &
 \qquad {\rm and} \,\qquad     \v Y(\tau,\s) =\v Y(\tau)~, \\ \nonumber
&\v{P} (\tau,\s) = \v{P}_0(\tau) + \v{P}_+(\tau)\, e^{i m \s} + \v{P}_-(\tau)\, e^{-i m \s} &
 \qquad {\rm and} \,\qquad	\v P_Y(\tau,\s) =\v P_Y(\tau) \, ,
\eea
where the positive integer $m \geq 0$ denotes the mode number of the single-mode ansatz.
Due to $\v Y'(\tau,\s)=0$, the spherical part of the Hamiltonian \eqref{eq:AdSS_EnSq&Ham} is only given by the first term in \eqref{eq:AdSS_SHam}, rendering the Hamiltonian polynomial in phase space variables. In particular, we are left with particle dynamics on $S^5$ and \eqref{eq:AdSS_SHam} coincides with the $SO(6)$ Casimir number calculated from \eqref{eq:AdSS_SO6Rots},
\be\label{eq:HRed_SO6Casimir}
{\cal M}_S^2=(1+\v{Y}^2/4)\v{P}_Y^2=\frac{1}{2}\,L_{i\,j}L_{i\,j}+L_{i\,M+1}L_{i\,M+1}~.
\ee

Since the AdS string dynamics is nonlinear, an excitation of one non zero-mode, in general, excites other modes
and we have to find conditions, which preserve the single-mode ansatz \eqref{eq:HRed_OMAnsatz} in dynamics. In addition the constraints in \eqref{eq:SSST_Constr1} should be satisfied, that is  
the Hamiltonian density  should be  $\s$-independent and the level matching density \eqref{eq:AdSS_LevelMatch} should vanish. 
In the Hamilton equations, which follow from \eqref{eq:AdSS_EnSq&Ham}, scalar combinations as
$\v{X}^2$, $\v{P}^2$, $\v{P}\cdot\v{X}$, {\it etc.}
play the role of coefficients of the vectors $\v{P},\,$ $\,\v{X}$ and their derivatives.
One can then show that the vanishing of the level matching density 
and the stability of the single-mode ansatz in dynamics require
these scalar combinations to be $\s$-independent.
By this, the Hamiltonian density also becomes  $\s$-independent. Explicitly, we find the following conditions on the excited modes
\bea\label{eq:HRed_ComplConstr}
&\v{X}_\pm\cdot \v{P}_0=\v{P}_\pm\cdot \v{X}_0=0 ~,
	\quad \v{X}_\pm\cdot \v{X}_0=\v{P}_\pm\cdot \v{P}_0=0~,\\ \nonumber
&\v{X}_\pm^2=\v{P}_\pm^2=0~,
	\quad \v{P}_\pm\cdot \v{X}_\pm=0~,
	\quad \v{P}_+\cdot \v{X}_- -\v{P}_-\cdot \v{X}_+=0~.
\eea

It is clear from these conditions that the zero-mode vectors $\v P_0$, $\v X_0$ may be considered as unconstrained. Then, 
introducing real and imaginary parts of the non zero-modes, $\,\,\v{X}_\pm=\frac{1}{\sqrt 2}(\v{X}_{\rm re}\pm i\v{X}_{\rm im})$ and
$\,\v{P}_\pm=\frac{1}{\sqrt 2}(\v{P}_{\rm re}\pm i\v{P}_{\rm im})$,
one sees that they  lie in the plane orthogonal to the plane spanned by the zero-mode vectors $\v P_0$, $\v X_0$.\footnote{For any solution of the classical equations of motion this zero-mode plane turns out to be defined by the boosts and, therefore, has a $\tau$-independent orientation in $\Reals^4$, as is shown below.}
In addition, the four non zero-mode vectors $\v X_{\rm re}$, $\v X_{\rm im}$, $\v P_{\rm re}$, $\v P_{\rm im}$ satisfy the constraints
\bea\label{eq:HRed_RealConstr}
&\v{X}_{\rm re}^2=\v{X}_{\rm im}^2~, \quad ~~~\v{X}_{\rm re}\cdot\v{X}_{\rm im}=0~, \qquad
&\v{P}_{\rm re}^2=\v{P}_{\rm im}^2~, \quad \v{P}_{\rm re}\cdot\v{P}_{\rm im}=0~,\\
\nonumber
&\v{X}_{\rm re}\cdot\v{P}_{\rm im}=0~, \quad \v{P}_{\rm re}\cdot\v{X}_{\rm im}=0~, \qquad &\v{P}_{\rm re}\cdot\v{X}_{\rm re}=\v{P}_{\rm im}\cdot\v{X}_{\rm im}~.
\eea
Since we are considering $AdS_5$, the non zero-mode vectors can be parameterized in the form
\be\label{eq:HRed_DefPQ}
\v{X}_{\rm re}=\frac{Q}{\sqrt 2}\,\v{\bf e}_{\rm re}~, \quad  \v{P}_{\rm re}=\frac{P}{\sqrt 2}\,\v{\bf e}_{\rm re}~,\qquad
\v{X}_{\rm im}=\pm\frac{Q}{\sqrt 2}\,\v{\bf e}_{\rm im}~, \quad \v{P}_{\rm im}=\pm\frac{P}{\sqrt 2}\,\v{\bf e}_{\rm im}~,
\ee
where $\v{\bf e}_{\rm re}$ and $\v{\bf e}_{\rm im}$ are orthonormal vectors with the standard relative orientation. One can always choose  
$\v{\bf e}_{\rm re}$ to be in the direction of  $\v{X}_{\rm re}$ which implies $Q \geq 0$. 
Then, there is no restriction on the sign of $P$,  i.e. $P\in \Reals^1$.
Note that the two solutions for $\v{X}_{\rm im}$ and $\v{P}_{\rm im}$ are related to each other by the reflection $\s\mapsto -\s$ which is a symmetry of the string action. 
In what follows we choose for definiteness the solution with all positive signs. 

Inserting the parametrization \eqref{eq:HRed_DefPQ} in the ansatz \eqref{eq:HRed_OMAnsatz}, we find
\be\label{eq:HRed_RedOMAnsatz}
\v{X} (\tau,\s) = \v{X}_0(\tau) + Q(\tau)\left[\v{\bf e}_{\rm re} \cos m \s + \v{\bf e}_{\rm im}\sin m \s\right]~,
\ee
and a similar expression for $\v{P} (\tau,\s)$.
The reduction of the AdS part of the initial canonical symplectic form to the single-mode ansatz \eqref{eq:HRed_OMAnsatz}  yields
\be\label{eq:HRed_RedSymplForm1}
\Omega=\int_0^{2\pi}\frac{\mbox{d}\s}{2\pi}\, \mbox{d}\v{P}(\s)\wedge \mbox{d}\v{X}(\s)= \mbox{d}\v{P}_0\wedge \mbox{d}\v{X}_0+
\mbox{d}\v{P}_{\rm re}\wedge \mbox{d}\v{X}_{\rm re}+\mbox{d}\v{P}_{\rm im}\wedge \mbox{d}\v{X}_{\rm im}~.
\ee
Using then the parametrization \eqref{eq:HRed_DefPQ} and the identities
$\,\mbox{d}\v{\bf e}_{\rm re} \wedge \mbox{d}\v{\bf e}_{\rm re}=\mbox{d}\v{\bf e}_{\rm im}\wedge \mbox{d}\v{\bf e}_{\rm im}=0$ as well as
$\,\,\v{\bf e}_{\rm re}\cdot \mbox{d}\v{\bf e}_{\rm re}=\v{\bf e}_{\rm im}\cdot \mbox{d}\v{\bf e}_{\rm im}=0$,  we  find the reduced symplectic form
\be\label{eq:HRed_RedSymplForm2}
\Omega=\mbox{d}\v{P}_0\wedge \mbox{d}\v{X}_0+\mbox{d}P\wedge \mbox{d}Q~.
\ee
It is independent of the orientation of the vector $\v{\bf e}_{\rm re}$ because the non zero-mode parts of $\v{X} (\tau,\s)$ and $\v{P} (\tau,\s)$ are collinear. This independence  may also be  interpreted as the residual gauge symmetry $\s\mapsto \s+f(\tau)$ of the  
single-mode ansatz as the level matching constraint is satisfied. 
Even including all the other excitations, one can always perform a canonical transformation which for the vectors $\v{X} (\tau,\s)$ and $\v{P} (\tau,\s)$ takes the form of a rotation in the non zero-mode plane, and change the orientation of  $\v{\bf e}_{\rm re}$ arbitrarily.

Thus, the Hamiltonian reduction of the AdS part leads to a ten-dimensional phase space with the canonical coordinates
$(\v{P}_0,\,\v{X}_0,\,P,\,Q)$, where the pair $(P,\,Q)$ lives on the half-plane $Q\geq 0$.

Calculating now the string energy square and the symmetry generators, where
the boosts are treated as 4-vectors $\v J_{0'}\equiv J_{a\,0'},$  $\,\v J_0 \equiv J_{a\,0}$,
from \eqref{eq:AdSS_EnSq&Ham} and \eqref{eq:AdSS_SO24Rots}-\eqref{eq:AdSS_SO24Boosts} we obtain
(recall $\omega=E/\sqrt{\lambda}$)
\be\label{eq:HRed_OMEnSquared}
E^2=2\sqrt{\l}\,H=\left(1+\v{X}_0^2+Q^2\right)\left(\v{P}^{\,2}_0+P^2+(\v{P}_0\cdot\v{X_0}+P Q)^2+{\cal M}_S^2+{\l}\,m^2 Q^2\right)~,
\ee
\be\label{1-mode AdS rotations}
J_{0'\,0} = E~,  \qquad \qquad \qquad J_{a\,b} = P_{0,a}\, X^b_0 - P_{0,b}\, X^a_0~,
\ee
\be\ba\label{eq:HRed_OMAdSBoosts}
&\v{J}_{0'} = \frac{E\, \v{X}_{0}}{\sqrt{1+\v{X}_0^2+Q^2}}\cos(\om\,\tau)-
\sqrt{1+\v{X}_0^2+Q^2}\,\,\v{P}_0\sin(\om\,\tau)~,& \\
&\v{J}_{0} =-\frac{E\, \v{X}_{0}}{\sqrt{1+\v{X}_0^2+Q^2}}\sin(\om\,\tau)-
\sqrt{1+\v{X}_0^2+Q^2}\,\,\v{P_{0}}\,\cos(\om\,\tau)~.&
\ea\ee

From \eqref{eq:HRed_OMEnSquared} and \eqref{eq:HRed_OMAdSBoosts} follow the Poisson brackets
\be\label{eq:HRed_PB-EnSq-Boosts}
\{E^2,J_{a\,0'}\} =-2E\,J_{a\,0}~,  \qquad \{E^2,J_{a\,0}\} = 2E\,J_{a\,0'}~,
\ee
which are equivalent to a part of the $\mathfrak{so}(2,4)$ commutation relations.
The validity of the remaining algebra is straightforward.	

In contrast to the particle dynamics, the SO$(2,4)$ Casimir
\be\label{eq:HRed_SO24Casimir}
C = E^2 + \f{1}{2} J_{a\,b} J_{a\,b} - \v{J}_{0'}^{\,2} - \v{J}_0^{\,2}
\ee
resulting from \eqref{eq:HRed_OMEnSquared}-\eqref{eq:HRed_OMAdSBoosts} is not a number, but the following function on the phase space
\be\label{eq:HRed_OMSO24Casimir}
C=(P+\tilde D\,Q)^2+(1 + Q^2) ({\cal M}^2_{S}+\l m^2\,Q^2 )~,
\ee
where $\tilde D=P\,Q+\v{P}_0\cdot\v{X}_0$ is the generator of dilatations in the ten-dimensional phase space.

Note that for vanishing zero-modes the SO$(2,4)$ Casimir and the energy square \eqref{eq:HRed_OMEnSquared} coincide.

In quantum theory we will be interested in the spectrum of the corresponding operators corrected by vacuum fluctuations and possible supersymmetric corrections.

\subsection{Integrability of single-mode strings} \label{subsec:IOMS}

The Hamilton function of the single-mode string defined by \eqref{eq:HRed_OMEnSquared} has the structure
of a Hamiltonian of a non relativistic particle. The kinetic part, given by the terms quadratic in momenta,
corresponds to a free-particle on the five dimensional unit half-sphere\footnote{More precisely,
it is a quarter of the sphere, since $Q\geq 0$}.
The term ${\cal M}_S^2$ in the potential corresponds to the SO(6) Casimir and can be treated as a constant
mass-square parameter. The potential contains terms quadratic and quartic in ($\v{X}_0,\, Q)$.
Formally taking $m=0$ in  \eqref{eq:HRed_OMEnSquared} one only has the quadratic terms, which
corresponds to the Higgs-potential on a sphere discussed in \cite{Higgs:1978yy}.
This model corresponds to the $AdS_6$ particle with mass ${\cal M}_S$ in the static gauge \cite{Dorn:2010wt, Jorjadze:2012jk} and it is exactly solvable in terms of trigonometric functions.
Below we show that the quartic terms do not destroy the integrability of the system, however the solution has a more
complicated structure. To show the integrability, we introduce a canonical map to new coordinates, which allows to separate
the non zero-mode variables and express the Casimir \eqref{eq:HRed_OMSO24Casimir} only through them.

Taking into account the form of \eqref{eq:HRed_OMSO24Casimir}, we introduce a transformation of the phase space coordinates
$(\v{P}_0,\,\v{X}_0,\,P,\,Q)\leftrightarrow (\v{p}_0,\,\v{x}_0,\,p,\,q)$ in the following form
\be\label{eq:IOMS_CanTrafo}
\v{P}_{0}=\frac{\v{p}_0}{\sqrt{1+q^2}}~,   \quad  \v{X}_0=\v{x}_0 \sqrt{1+q^2}~,
\quad P=p-\frac{q(\v{p}_0 \cdot\v{x}_0)}{1+q^2}~, \quad  Q=q~.
\ee
It is straightforward to check that \eqref{eq:IOMS_CanTrafo} is a canonical transformation
\be\label{eq:IOMS_CanonSymplForm}
	\mbox{d}\v{P}_0\wedge \mbox{d}\v{X}_0+\mbox{d}P\wedge \mbox{d}Q=\mbox{d}\v{p}_0\wedge \mbox{d}\v{x}_0 + \mbox{d}p\wedge \mbox{d}q~,
\ee
related to the change of coordinates $(\v X_0, Q)\leftrightarrow (\v{x}_0, q)$.

In the new variables, the SO(2,4) dynamical integrals \eqref{eq:HRed_OMEnSquared}-\eqref{eq:HRed_OMAdSBoosts}
take the form of a free massive particle in $AdS_5$
\be\label{eq:IOMS_EnSquared}
	E^2 = 2 \sqrt{\l}\,H = \left(1+\v{x}_0^{\,2}\right)\left(\v{p}_0^{\,2}+(\v{p}_0\cdot\v{x}_0)^2 +{\cal M}^2\right)~,
\ee
\be\label{eq:IOMS_OMAdSRots}
	J_{0'\,0} = E~,  \qquad \qquad \qquad J_{a\,b} = p_{0,a}\, x^b_0 - p_{0,b}\, x^a_0~,
\ee
\be\ba\label{eq:IOMS_OMAdSBoosts}
	&\v{J}_{0'} = \frac{E\, \v{x}_0}{\sqrt{1+\v{x}_0^{\,2}}}\cos(\om\,\tau)-
\sqrt{1+\v{x}_0^{\,2}}\,\,\v{p}_0\sin(\om\,\tau)~,&
\\
	&\v{J}_{0} =-\frac{E\, \v{x}_0}{\sqrt{1+\v{x}_0^{\,2}}}\sin(\om\,\tau)- \sqrt{1+\v{x}_0^{\,2}}\,\,\v{p}_0\,\cos(\om\,\tau)~,&
\ea\ee
and the mass-square term depends only on the non-zero mode phase space variables $(p,\, q)$
\be\label{eq:IOMS_MassSquared}
	{\cal M}^2=(1+q^2)\left[p^2+p^2 q^2+{\cal M}_S^2+{\l}\,m^2 q^2\right]~.
\ee
This term coincides with the SO(2,4) Casimir calculated from \eqref{eq:IOMS_EnSquared}-\eqref{eq:IOMS_OMAdSBoosts}
and, therefore, it is $\tau$-independent, like the SO(6) Casimir ${\cal M}_S^2$.

It follows from \eqref{eq:IOMS_OMAdSBoosts} that
$\sqrt{1+\v{x}_0^{\,2}}\left[\v{J}_{0'}\cos(\om\,\tau)-\v{J}_{0}\sin(\om\,\tau)\right]=E\, \v{x}_0$.
The squaring of this relation, defines the scalar $\v{x}_0^{\,2}$ in terms of the boosts and $\tau$-dependent
trigonometric functions.
The pair $(\v{x}_0,\,\v{p}_0)$ is then obtained from \eqref{eq:IOMS_OMAdSBoosts} algebraically,
as a solution of the linear system, with given $\tau$-dependent coefficients.

Note that the integrals of motion \eqref{eq:IOMS_OMAdSRots}-\eqref{eq:IOMS_OMAdSBoosts} are related by
$E\,J_{a\,b}=J_{a\,0'}\,J_{b\,0}-J_{b\,0'}\,J_{a\,0}$. This, relation together with \eqref{eq:HRed_SO24Casimir},
allows to express $E$ and $J_{ab}$ through the boosts \cite{Dorn:2005ja}.
Thus, the pair $(\v{x}_0,\,\v{p}_0)$ is parameterized by $(\v J_{0'}$, $\v J_0$) and the Casimir number ${\cal M}^2$ only.

The dynamics of the $(p,\,q)$ pair is governed by the Hamiltonian
\be\label{eq:IOMS_Hq}
	H_m=\frac{1+\v{x}_0^{\,2}(\tau)}{2\sqrt\l}\,{\cal M}^2~,
\ee
where $\v{x}_0^{\,2}(\tau)$ has been calculated above and ${\cal M}^2$ is given by \eqref{eq:IOMS_MassSquared}.
Eliminating the momentum variable $p$ from the Hamilton equation for $\dot q$ by
solving \eqref{eq:IOMS_MassSquared} for $p$, one finds
\be\label{eq:IOMS_qdot}
	\pm\sqrt\l\,\,\dot q=(1+\v{x}_0^{\,2}(\tau))(1+q^2)\sqrt{{\cal M}^2-(1+q^2)({\cal M}_S^2+{\l}\,m^2 q^2)}~.
\ee
This equation is separable in $(q,\,\tau)$ and can be integrated directly, therefore proving the asserted classical integrability of the single-mode system. The $q$-integration however
results in an elliptic integral of third kind and the solutions can not be written in terms of elementary functions.

To find solutions at large $\l$ and vanishing zero-modes, we rescale the phase space variables
\be\label{eq:IOMS_pqRescale}
	q\mapsto\frac{q}{\l^{1/4}\sqrt{m}}~, \qquad  p\mapsto\l^{1/4}\sqrt{m}\,p~,
\ee
and obtain, in the leading order, the oscillator Hamiltonian
\be\label{eq:IOMS_HqExpand}
	H_m=\,\frac{m}{2}\,(p^2+q^2)+{\cal O}(\l^{-1/2}) ~,
\ee
which provides time oscillations of the $q$ coordinate. Since this coordinate is non-negative and it describes
the length of the vectors $\v{X}_{\rm re}$ and $\v{X}_{\rm im}$ in the plane of non zero-modes, one gets a circular string with an oscillating radius.

\begin{figure}[h!] 
	\centering
	\begin{tikzpicture}
	\node[above right] (img) at (0,0) {\includegraphics[width=0.12\textwidth]{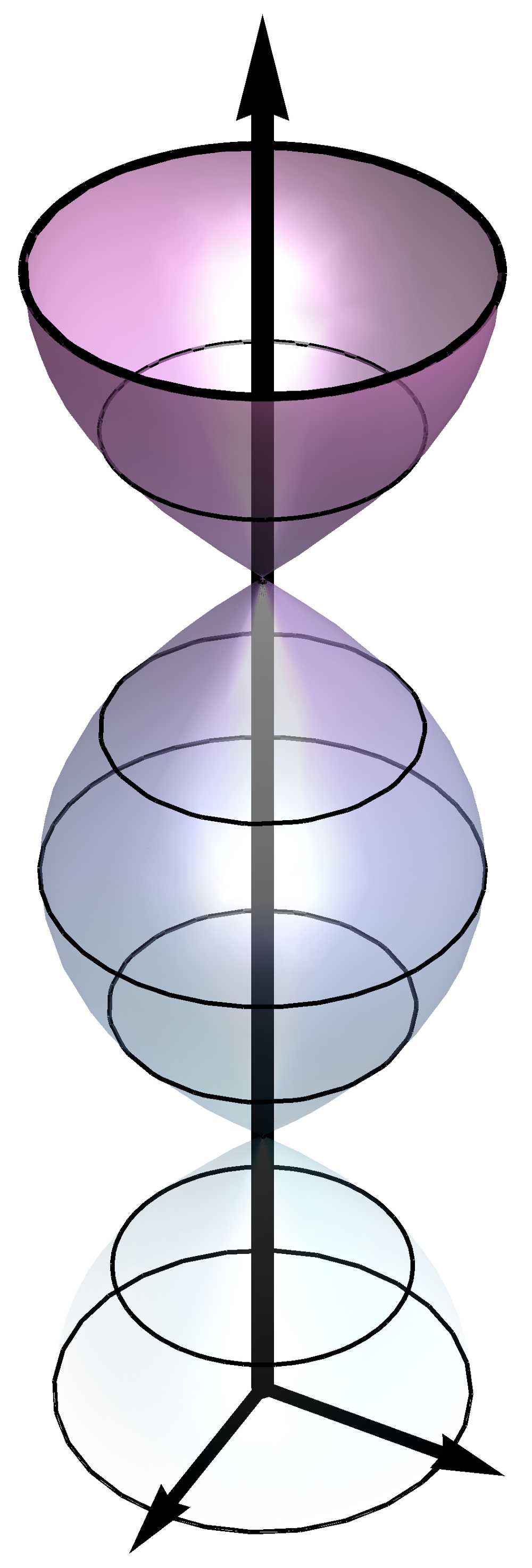}};
	\node at (55pt,185pt) {\footnotesize $\vXz, \vPz \in \Reals^2$};
	\node at (15pt,0pt) {\footnotesize $\vXr, \vPr \in \Reals$};
	\node at (90pt,13pt) {\footnotesize $\vXi, \vPi \in \Reals$};
	\node at (-2pt,187pt) {\footnotesize $\Reals^4$};
     \draw (-10pt,179pt) -- (6pt,179pt);
	\draw (6pt,193pt) -- (6pt,179pt);
	\node at (-2pt,150pt) {\scriptsize $\tau_0$};
	\node at (3pt,126pt) {\scriptsize $\tau < \tau_0$};
	\end{tikzpicture}
					\caption{Single-mode solution, basically a pulsating string, depicted in the four spatial dimensions of $AdS_5$, with the zero mode plane projected to a line. The string area is traced out up to some world sheet time $\tau_0$, thin lines showing string configurations for $\tau < \tau_0$. We neglected that the zero-modes oscillate themselves.}
					\label{fig:PulsatingString}
	\end{figure}
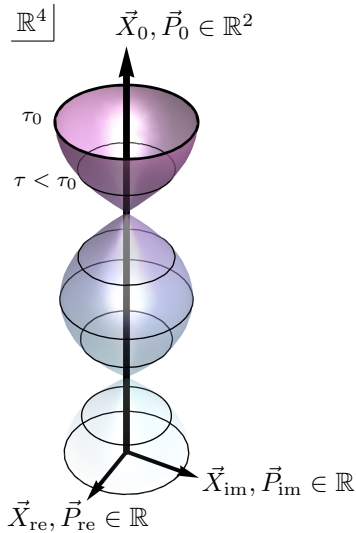

This type of string solutions are called pulsating strings and are depicted in \figref{fig:PulsatingString}.
In AdS spaces they were obtained first for $AdS_3$ \cite{deVega:1994yz} and then they have been applied in the context of AdS/CFT in \cite{Minahan:2002rc}, where however in both studies zero-modes have been disregarded, $X^k_0 = P_{0,k} = 0$. The virtue of the presented single-mode solution therefore lies in the allowance for zero-modes and their exact solution as massive particles in $AdS_5 \times S^5$.

Finally, note that by \eqref{eq:IOMS_pqRescale} we have ${\cal M}^2 \propto \sqrt{\l}\,m$. Taking then the rescaling $\v x_0 \mapsto \v x_0/\sqrt{{\cal M}}$, $\v p_0 \mapsto \sqrt{{\cal M}}\,\v p_0$, the leading order zero-modes Hamiltonian \eqref{eq:IOMS_EnSquared} becomes a harmonic oscillator
\be\label{E2leading}
E^2 = {\cal M}^2 + {\cal M}(\v{p}_0^{\,2}+\v{x}_0^{\,2})+\cdots~,
\ee
as has been observed before in \cite{Passerini:2010xc}. In particular, the zero modes scale differently in $\l\,m^2$ than the non zero-modes and are not free particles, not even in leading order. 


\section{Calculation of the energy spectrum} \label{sec:EnCalc}

Here we quantize the non zero-mode sector, construct the Casimir operator ${\cal M}^2$, heuristically taking into account corrections due to fermions, and calculate its spectrum perturbatively up to next-to leading order.
The energy spectrum of the pulsating string is then extracted similarly to the $AdS_5$ particle.

\subsection{Quantization of non zero-mode sector} \label{subsec:QNZM}

The energy spectrum of the $AdS_5$ scalar particle with Casimir number 
${\cal M}^2$ is of harmonic oscillator type (see 
e.g.~\cite{Dorn:2005ja,Dorn:2010wt, Jorjadze:2012jk}),
\be\label{eq:QNZM_MinEnergy}
E_{(N,{{\cal M}^2})}=2+\sqrt{\mathcal{M}^2+4} + N\,, \qquad N\in \mathbb{N}~.
\ee
With help of the canonical transformation established in the last section 
the energy spectrum of the pulsating string then follows from the spectrum of the
operator  ${\cal M}^2$ defined by \eqref{eq:IOMS_MassSquared} depending only on the
string-modes and the ${\cal M}_S^2$ of the center-of-mass motion on the $S^{5}$.
Expanding \eqref{eq:QNZM_MinEnergy} in powers of $1/{\cal M}$, one gets for the ground state 
$E_{(0,{{\cal M}^2})}\approx \mathcal{M}+ 2\,$, which is perfectly consistent with what follows from the leading order zero-mode Hamiltonian \eqref{E2leading}, with the constant $2$ appearing because the $AdS_5$ zero-mode vectors are four-dimensional. 
Note then that the eigenvalues of the SO(6) Casimir number are $J(J+4)$ with non-negative integer $J$ and in the bosonic case
we replace ${\cal M}_S^2$ in \eqref{eq:IOMS_MassSquared} by this value. There is an 
immediate puzzle arising if we consider the single-mode string in its groundstate
in the $AdS_{5}$ sector, i.e., a point particle rotating on the $S^{5}$ only, ${\cal M}^2={\cal M}_S^2=J(J+4)$, we obtain $E_{(0,J(J+4))}=J +4$. However, $E_{(0,\mathcal{M}^2_S)}$ should be equal to $J$
by comparison to the plane-wave superstring in light-cone gauge with vanishing
$E-J$ for the groundstate \cite{Berenstein:2002jq}. This state is dual to the protected gauge theory operator $\Tr(Z^{J})$. 

The resolution of this puzzle lies in our ignorance
of fermions. For the point particle, the inclusion of the fermionic degrees of freedom
changes the dependence of ${\cal M}_S^2$ on $J$ to ${\cal M}_S^2=J(J-4)$ giving thus the expected result \cite{Metsaev:1999gz}. For string excitations one would expect in particular, that the eight physical fermionic zero-modes included in the reduced Hamiltonian \eqref{eq:IOMS_EnSquared} would produce in the leading order the additional constant $-4$, changing the constant $2$ in 
the ground state energy to $-2$. Based on this and the
comparison to the plane-wave light-cone groundstate, we thus conclude that the large $\cal M$ expansion of the energy spectrum is given by
\be\label{eq:QNZM_MinEnergy2}
E_{(N,{{\cal M}^2})}=\mathcal{M}-2+\frac{2}{\mathcal{M}} + N+\cdots\,, \qquad N\in \mathbb{N}~,
\ee
where $\cal M$ may acquire additional $1/\lambda^{1/4}$ corrections due to fermions. 

Of course, this is only a heuristic argument at this point and should be backed up through an analysis of the full superstring. 

Let us now turn to the spectrum ${\cal M}^2$ and quantize the string-mode 
sector.  For this one has to choose
either Dirichlet or Neumann boundary conditions for  wave functions $\psi(q)$ 
which are defined on the half-line $q>0$ at $q=0$.
This ambiguity is related to the non-selfadjointness of the momentum operator $p$ on the half-line $q\geq 0$.

To avoid this ambiguity, we use the quantization scheme proposed in Section 2 for the
static gauge approach. In our case, this assumes a canonical quantization of the modes
$(\v P_0,\, \v X_0)$, $(\v P_\pm,\, \v X_\pm)$
and restriction of the Hilbert space by the quantum version of the constraints \eqref{eq:HRed_ComplConstr}.
For the analysis of these constraints it is convenient to introduce two 8-dimensional vectors constructed by
the pairs ($\v X_{\rm re},$ $\v X_{\rm im}$) and ($\v P_{\rm re},$ $\v P_{\rm im}$),
respectively. It has to be noted that, on the constraint surface \eqref{eq:HRed_RealConstr},
these two 8-vectors are parallel to each other, which can be easily seen from the
representation \eqref{eq:HRed_DefPQ}.
The Hamiltonian reduction in the non zero-mode sector is then equivalent to the gauging of the
16 dimensional phase space with coordinates {
($\v P_{\rm re},$ $\v P_{\rm im}$; $\v X_{\rm re},$ $\v X_{\rm im}$)} by the group SO(8).
As a result, the non zero-mode part of the physical wave functions
depends only on the SO(8) scalar $q=\sqrt{\v {X}^2_{\rm re}+ \v {X}^2_{\rm im}}$.

One can generalize this scheme, introducing two ${{N_q}}$-dimensional vectors ($\v p$, $\,\v q$), which
replace the pair ($p$, $\, q$) in \eqref{eq:IOMS_MassSquared} as follows
\be\label{eq:QNZM_MassSquared}
{\cal M}^2=(1+\v q^{\,2})\left[\v p^{\,2}+(\v p \cdot \v q)^{2}+{\cal M}^2_S+{\l}\,m^2 \v q^{\,2}\right]~.
\ee
The gauging of the phase space ($\v p$, $\v q$)
by the SO(${{N_q}})$ group leads to the half-plane $(p,\, q)$ where $q$ is the length of the vector $\v q$ and
$p=(\v p\cdot\v q)/q$. The Casimir \eqref{eq:QNZM_MassSquared} is gauge invariant and it obviously reduces to \eqref{eq:IOMS_MassSquared}.
Quantum theory on the half-plane $(p,\, q)$ is then identified
with the ${N_q}$-dimensional case, restricted
to the wave functions $\psi(q)$.

Note that solving the constraints \eqref{eq:HRed_ComplConstr} partially, one gets similar systems
with ${N_q}<8$. For example, taking into account that non zero-modes are orthogonal to zero-modes,
the pairs ($\v X_{\rm re},$ $\v X_{\rm im}$) and ($\v P_{\rm re},$ $\v P_{\rm im}$) can be treated as 4-vectors
and one has to perform the SO(4) gauging.

Using the rescaling \eqref{eq:IOMS_pqRescale} for ${N_q}$-dimensional vectors $\v p$ and $\v q$,
from \eqref{eq:QNZM_MassSquared}
we obtain
\be\label{eq:QNZM_MSqExpand}
{\cal M}^2=\sqrt\l\,m\left(\v p^{\,2}+\v q^{\,2}\right)+\v p^{\,2}\,\v q^{\,2}+(\v p \cdot \v q)^{2}+(\v q^{\,2})^2+{\cal M}^2_S+{\cal O}(\l^{-1/2})~.
\ee

The ${{N_q}}$-dimensional oscillator Hamiltonian
\be\label{eq:QNZM_HqLead}
h=\frac{1}{2}(\v p^{\,2}+\v q^{\,2})
\ee
has eigenfunctions with zero angular momentum at the even levels $2n$ only.
We denote these states by $|n\rangle$. They are obtained by the action of the operators $(\v a^{\,\dag\,\,2})^n$
on the vacuum state. Hence, these wave functions are even in $q$ and they correspond
to the Neumann boundary condition at $q=0$.
In the appendix we present an algebraic construction of these states and calculate
mean values of scalar operators.
Since the oscillator Hamiltonian \eqref{eq:QNZM_HqLead} describes ${\cal M}^2$ in the leading order,
we can use these calculations to find the spectrum of  $\mathcal{M}^2$ perturbatively.

The construction of the operator ${\cal M}^2$ by its classical expression \eqref{eq:QNZM_MSqExpand} contains operator ordering ambiguities.
It has to be noted that for a given Hermitian ordering the mean values of a scalar operator essentially depends on the dimension ${{N_q}}$ (see \eqref{eq:SONq_MeanVals1}-\eqref{eq:SONq_MeanVals2}
in the appendix).

Hence, to fix the operator ${\cal M}^2$, we first rewrite \eqref{eq:QNZM_MSqExpand} in terms of creation-annihilation variables, keeping its scalar structure. For this purpose we introduce complex scalar functions
\be\label{eq:QNZM_AAdag}
A=\frac{1}{2}(\v p^{\,2}-\v q^{\,2})-i\v p\cdot\v q~, \quad
A^\dag=\frac{1}{2}(\v p^{\,2}-\v q^{\,2})+i\v p\cdot\v q~,
\ee
which correspond to the operators $\v a^{\,2}$ and $\v a^{\,\dag\,\,2}$, respectively (see \eqref{eq:SONq_ONqScalars}).

The Casimir \eqref{eq:QNZM_MSqExpand} in the new variables becomes
\be\label{eq:QNZM_MSq-hAAdag}
{\cal M}^2=2\sqrt\l\,m\,h+2\,h^2+\f{1}{2}\,A^\dag\,A-\f{1}{4}(A^{\dag\,2} +A^2)
-h(A^\dag+A)+{\cal M}^2_S + \ord{\l^{-1/2}}~.
\ee
Choosing a normal ordering with ${\colon\!}h{\colon\!}=h-{{N_q}}/2$ (see \appref{app:SONq}), one has to make the replacement
$$h^2\mapsto {{\colon\!}h^2{\colon\!}=\;}({\colon\!}h{\colon\!})^2-{\colon\!}h{\colon\!}~.$$ 
With the help of 
\be\label{eq:QNZM_MeanValues}
\langle n|{\colon\!}h{\colon\!}|n\rangle =2\,n~, \qquad \langle n|A^\dag\,A|n\rangle=4\,n^2+2({{N_q}}-2) n~,
\ee
the mean values of the operator ${\cal M}^2$ become
\be\label{eq:QNZM_MSqSpectrum1}
\langle n|{\cal M}^2|n\rangle=4\,{\sqrt{\l}}\,m\, n +10\,n^2 +({{N_q}}-6)n+{\cal M}^2_S
+ \ord{\l^{-1/2}}~.
\ee
Let us now argue for a way to mimick the effect of supersymmetry whose implementation via
the quantum symmetry algebra should hopefully fix (some) of the ordering ambiguities
encountered in our bosonic string analysis.  Recall that the bosons give rise to the value $N_{q}=8$ in the above equation \eqref{eq:QNZM_MSqSpectrum1}
 related to the
8 dimensional phase space of the four transverse $AdS_{5}$ dimensions of our static gauge. 
In the spirit of effective ``negative dimensions'' for fermions we 
argue that taking them into account will
lead to an effective value of ${{N_q}}=0$ in the final 
expressions for the spectrum. This approach is certainly true for the super-harmonic oscillator
present at the leading perturbative order and we will implement it at the orders as well.

Taking  ${{N_q}}=0$ in \eqref{eq:QNZM_MSqSpectrum1} we obtain
\be\label{eq:QNZM_MSqSpectrum2}
\langle n|{\cal M}^2|n\rangle=4\,{\sqrt{\l}}\,m\,n + 10\,n^2-6n+{\cal M}^2_S~.
\ee
Equation \eqref{eq:QNZM_MinEnergy2} together with ${\cal M}^2_S=J(J+4)$ then leads to  the following energy spectrum
\be\label{eq:QNZM_EnSpectrum}
E_{N,m,n,J}=2\,\l^{1/4}\sqrt{m\,n} +N-2+\frac{10\,n^2-6\,n+4+J(J+4)}{4\,\l^{1/4}\sqrt{m \,n}}
+ \ord{\l^{-3/4}}~,
\ee
with $N$ being the excitation of the $AdS_{5}$ particle degrees of freedom of \eqref{eq:QNZM_MinEnergy2}. In fact we expect that the inclusion of fermions would change ${\cal M}^2_S$ to $J^2$ as it follows from the flat space spectrum
and is supported by numerical studies of the TBA equations.
For the lowest stringy excitation of $N=0$, $m=1$, $n=1$ and $J=0$ conjecturally dual to the
Konishi operator this reduces to
\be\label{eq:QNZM_En11}
E_{0,1,1,0}=2\,\l^{1/4}-2+\f{2}{\l^{1/4}}+ \ord{\l^{-3/4}}~.
\ee
It is now important to ask to what extent we can trust the level truncation scheme that
we have pursued by restricting the quantum dynamics of the string to a single non-zero mode excitation. Clearly at some order in perturbation theory this minisuperspace
approach fails to provide the correct energy of the studied string excitation as 
the suppressed non-zero modes would contribute in intermediate states in perturbation theory
of the full model. The 
minisuperspace quantization for the lowest mode string with $n=m=1$ will yield a better
approximation to the energy spectrum than for higher excitations $n,m>1$ as here the fluctuation
of lower modes will contribute earlier. 
In the following section we will quantify this statement in some detail. Indeed the
intuitive expectation is confirmed, namely that 
the result (\ref{eq:QNZM_EnSpectrum}) at order $\mathcal{O}(\lambda^{-1/4})$ can only be 
trusted for the $m=n=1$ case - modulo the discussed ordering ambiguity.

\subsection{Decoupling of other modes} \label{subsec:DoOM}
To investigate {to which order and for which states our result \eqref{eq:QNZM_EnSpectrum} can be trusted we have to return again to} the full string action, i.e., consider again all modes, not only the zero-modes and $m$th $AdS_5$ mode, and then see at which order the modes disregarded in \eqref{eq:HRed_OMAnsatz} could contribute to the energy. Generally, for this one should investigate the full type IIB superstring. However, since a suitable prescription of static gauge for the Green-Schwarz superstring is lacking, we follow the line of thought so far and constrain our investigation to the bosonic subsector, where again we will argue about the effects of supersymmetry heuristically. Hence, we return to \eqref{eq:AdSS_EnSq&Ham}.

	For the spatial $AdS_5$ phase space variables as well as the $S^5$ non zero-modes we then employ the rescaling suggested in \cite{Passerini:2010xc}, see also the comment beneath \figref{fig:PulsatingString}, but we do not rescale the $S^5$ zero-modes. Hence, the rescaling of the spatial $AdS_5$ and $S^5$ phase space variables reads
	\begin{align}
		\vec{X}(\tau,\s) &\mapsto \l^{-1/4} \vec{x}(\tau,\s) + \l^{-1/8} \vec{x}_0 (\tau) ~,& \quad
		\vec{P}_X(\tau,\s) &\mapsto \l^{1/4} \vec{p}_x(\tau,\s) + \l^{1/8} \vec{p}_{x,0} (\tau) ~,&
		\label{eq:DoOM_RescX} \\
		\vec{Y}(\tau,\s) &\mapsto \l^{-1/4} \vec{y}(\tau,\s) + \vec{y}_0 (\tau) ~,& \quad
		\vec{P}_Y(\tau,\s) &\mapsto \l^{1/4} \vec{p}_y(\tau,\s) + \vec{p}_{y,0} (\tau) ~,&
		\label{eq:DoOM_RescY}
	\end{align}
	with $\vx(\tau,\s)$, $\vpx(\tau,\s)$, $\vec{y}(\tau,\s)$ and $\vec{p}_y(\tau,\s)$ comprising the respective $AdS_5$ and $S^5$ non zero-modes. It is important to note, that by this the
	expansion in powers of phase space variables is {\it
 not} the same as expansion in 't Hooft coupling $\l$. The Hamiltonian density (c.f. \eqref{eq:SSST_EnSquared}) then expands as 
	\begin{align}
		\calH 
		&= \ \f{1}{2}\left(\vpx^{\,2} + \vxpr^2 + \f{(1+\vyz^{\,2})^2}{4} \vpy^{\,2} + \f{4}{(1+\vyz^{\,2})^2} \vypr^{2}\right) 
			\nonumber\\
		&\ + \f{\l^{-1/4}}{2} \bigg\{\vpxz^{\,2} + \left(\vpx^{\,2} + \vxpr^2 + \f{(1+\vyz^{\,2})^2}{4} \vpy^{\,2}
			+ \f{4}{(1+\vyz^{\,2})^2} \vypr^{2}\right) \vxz^{\,2} + (\vpx\cdot\vxz)^2 - (\vxpr \cdot\vxz)^2 \
				\label{eq:DoOM_HamDens}\\
		& \qquad\quad\quad\ \   
		 + \frac{4\,\vy\cdot\vyz}{(1+\vyz^2)}
				\left(\f{(1+\vyz^{\,2})^2}{4} \vpy^{\,2}  - \f{4}{(1+\vyz^{\,2})^2} \vypr^{2} \right)\bigg\}
			\,+\, \ord{\l^{-3/8}}~,\nonumber
	\end{align}
	where we dropped terms linear in non zero-modes, e.g., the term $\l^{-1/8}\, \vpx\cdot\vpxz$,  as they obviously do not appear in the Hamiltonian $H = \f{E^2}{2\lll} = \int\!\sf{\dd \s}{2\pi}\calH$.
Nevertheless, \eqref{eq:DoOM_HamDens} looks alarming since at $\ord{\l^{-1/4}}$ in $H$ there will be several operators potentially contributing via second order perturbation theory to $\ord{\l^{-1/4}}$ in the energy $E$. Also, with the single-mode ansatz being a particle on $S^5$, the appearance of $\vyz$ at leading order but no $\vpyz$ seems unsatisfactory.

	These problems are overcome by a restricted canonical transformation to a new set of phase space variables, $(\vcpx, \vcx, \vcpxz, \vcxz, \vcpy, \vcy, \vcpyz, \vcyz)$, generated by a type II generating functional
	\be
		\mathcal{F}
			= \int \frac{\dd \s}{2 \pi}\left( \vcpxz\cdot\vxz + \vcpx\cdot\vx + \k\,(\vcpx\cdot\vxz)(\vxz\cdot\vx)
				+ \vcpyz\cdot\vyz + \f{2}{1+\vyz^{\,2}} \vcpy\cdot\vy \right)~,
	\ee
	with $\k \equiv \f{1}{2\,\l^{1/4}}(1-\f{3}{4\,\l^{1/4}}\vxz^2)$, yielding $\vcxz = \vxz$, $\vcyz = \vyz$ and
	\bea
		&\vcx = \vx + \k (\vx\cdot\vxz) \vxz\ \qquad \Leftrightarrow\qquad\
			\vx = \f{\vcx + \k (\vcxz^{\,2}\,\vcx - (\vcx\cdot\vcxz) \vcxz)}{1+\k\,\vcxz^{\,2}}~,& \nonumber\\
		AdS_5 :\qquad&\vpx = \vcpx + \k (\vcpx\cdot\vcxz) \vcxz \qquad\Leftrightarrow\qquad
			\vcpx = \f{\vpx + \k (\vxz^{\,2}\,\vpx - (\vpx\cdot\vxz) \vxz)}{1+\k\,\vxz^{\,2}}~,& \\
		&\vpxz = \vcpxz + \int \frac{\dd \s}{2 \pi} \left(\k\left(\vcpx(\vxz\cdot\vx)+(\vcpx\cdot\vxz)\vx\right)
			+ \f{\p \k}{\p \vxz} (\vcpx\cdot\vxz)(\vxz\cdot\vx)\right)~,& \nonumber\\[.3em]
		S^5:\ \ \qquad&\vy = \f{1+\vcyz^{\,2}}{2} \vcy~,\qquad \vpy = \f{2}{1+\vcyz^{\,2}} \vcpy~,
			\qquad \vpyz = \vcpyz - \f{2\,\vcyz}{(1+\vyz^{\,2})} \int\!\frac{\dd \s}{2 \pi}(\vcpy\cdot\vcy) ~.&
	\eea
	Notice that the new phase space variables differ from the old ones only by terms vanishing in our single-mode approximation {\eqref{eq:HRed_ComplConstr}, in particular $\vcpx\cdot\vcxz \propto \vpx\cdot\vxz = \vx \cdot \vxz = \vpy = \vy =0$}. Hence, the single-mode solution does not feel the canonical transformation and our result \eqref{eq:QNZM_EnSpectrum} persists.

	The new Hamiltonian density is then just the old one expressed in new phase space variables,
	\begin{align}
				\calH 
		&= \ \f{1}{2}\left(\vcpx^{\,2} + \vcxpr^2 + \vcpy^{\,2} + \vcypr^{2}\right) 
			\label{eq:DoOM_HamDens2}\\ 
		&\ + \f{\l^{-1/4}}{2} \bigg\{\vcpxz^{\,2}+\left(\vcpx^{\,2}+\vcxpr^2+\vcpy^{\,2}+\vcypr^{2}\right) \vcxz^{\,2}
			+ 2\,(\vcyz\cdot\vcy) \big(\vcpy^{\,2} - \vcypr^2\big)
			\bigg\} \,+\, \Ord{\l^{-3/8}}~, \nonumber
	\end{align}
	where again we ignored terms linear in non zero-modes.
	Now, in contrast to \eqref{eq:DoOM_HamDens}, the leading term in \eqref{eq:DoOM_HamDens2} is quadratic in non zero-mode phase space variables only and gives nothing but the leading flat space limit without the zero-modes,
	which feel the curvature of the background even at leading order.

	The first two terms of the second line are exactly the diagonal terms considered before determining the $AdS_5$ zero-modes to be harmonic oscillators in leading order, where for our solution $\vev{\vcpy^{\,2} + \vcypr^2}=0$. In particular, we cast off the off-diagonal terms in the second line of \eqref{eq:DoOM_HamDens}.\footnote{This illustrates that the $AdS_5$ part of the canonical transformation presented here is the analog of the unitary transformation found in \cite{Passerini:2010xc}.}

	The remaining $S^5$ term, the last term in the second line, has an odd power in non zero-mode phase space variables and {\it will} generally contribute to $\ord{\l^{-1/4}}$ in the energy via second order perturbation theory. 
	However, the terms will only give a contribution to the $\vcyz$ mass term, which for the $S^5$ non zero-modes in the ground state is expected to vanish due to supersymmetry. Hence, from this term and further $S^5$ terms at $\ord{\l^{-1/2}}$ in the Hamiltonian effectively only the massless particle on $S^5$ should survive.

	Therefore, we split the Hamiltonian into an unperturbed Hamiltonian and a perturbation as 
	\begin{align}
		&H = H_0 + \l^{-1/4}\,\d H = H_0 + \l^{-1/4}\,\d H_{-1/4} + \l^{-3/8}\,\d H_{-3/8}
			+ \l^{-1/2}\,\d H_{-1/2} + \ord{\l^{-5/8}} ~,\\
		&H_0 \equiv \int \f{\dd \s}{4 \pi} \left\{\left(\vcpx^{\,2} + \vcxpr^2 + \vcpy^{\,2} + \vcypr^{2}\right) + \f{\l^{-1/4}}{2} \left(\vcpxz^{\,2} + \left(\vcpx^{\,2} + \vcxpr^2 + \vcpy^{\,2} + \vcypr^{2}\right) \vcxz^{\,2}\right)\right\}~,\\
		&\d H_{-1/4} \equiv \int \f{\dd \s}{2 \pi} \left\{(\vcyz\cdot\vcy) \big(\vcpy^{\,2} - \vcypr^2\big)
			\right\}~, \label{eq:DoOM_dH14}\\
		&\d H_{-3/8} \equiv \int \f{\dd \s}{4 \pi} \Big\{\vcpx\cdot\vcxz\left(\vcpx\cdot\vcx + \vcpxz\cdot\vcxz\right)
			+ \vcx\cdot\vcxz \left(\vcpx^{\,2} + 2\,\vcxpr^2 \right) - 2(\vcx\cdot\vcxpr)(\vcxpr\cdot\vcxz)\Big\}~,
			\label{eq:DoOM_dH38}\\
		&\d H_{-1/2} \equiv \int \f{\dd \s}{4 \pi} \Big\{ \vcx^{\,2}\left(\vcpx^{\,2} + \vcxpr^2\right)
			+ \left(\vcpx\cdot \vcx+\vcpxz\cdot \vcxz\right)^2 + \vcpxz^{\,2}\,\vcxz^{\,2} - (\vcx\cdot \vcxpr)^2
			\label{eq:DoOM_dH12}\\
		& \qquad\qquad\qquad\quad\quad + (\vcpx\cdot\vcxz)(\vcpxz\cdot\vcx) + 3 (\vcpx\cdot\vcpxz)(\vcxz\cdot\vcx)
		   \,+\, \text{\{$S^5$ contributions\}}\Big\}~, \nonumber
	\end{align}
	where by the previous argument we did not spell out the plethora of $S^5$ terms in $\d H_{-1/2}$.
	{Let us} furthermore denote the eigenvalues of $H$ ($H_0$) of eigenstates $\ket{\psi}$ ($\ket{\psi^{(0)}}$) as $\calE_\psi=\f{E^2}{2\lll}$ ($\calE^{(0)}_\psi$).

	The terms in the first line of \eqref{eq:DoOM_dH12} then directly correspond to the terms present in \eqref{eq:HRed_OMEnSquared} when plugging in the single mode ansatz \eqref{eq:HRed_OMAnsatz}, where $\vcx\cdot\vcxpr = 0$ followed from the Virasoro constraints. On the other hand, the $AdS_5$ operators in the second line are expected to give no contribution to the energy at $\ord{\l^{-1/4}}$, since they are off-diagonal plus potentially a normal ordering constant, which ought to be canceled by supersymmetry.

	What is left is to discuss potential contribution from $\d H_{-3/8}$ \eqref{eq:DoOM_dH38}. Since the operators have an odd power in non zero-modes they will not contribute via first order perturbation theory. In second order perturbation theory \eqref{eq:DoOM_dH38} contributes as
	\be
		\calE^{(2)}_{\psi} \supset \sum_{\ket{\phi^{(0)}} \neq \ket{\psi^{(0)}}}
			\f{\big|\bra{\psi^{(0)}}\l^{-3/8} \d H_{-3/8} \ket{\phi^{(0)}}\big|^2}{\calE^{(0)}_{\psi} -
\calE^{(0)}_{\phi}} ~. \label{eq:DoOm_2ndOrdPert}
	\ee
	For states $\ket{\psi^{(0)}}$ with different level than $\ket{\psi^{(0)}}$ one has $\calE^{(0)}_{\psi} -
\calE^{(0)}_{\phi} \propto \l^0$ and the second order contribution will be of order $\ord{\l^{-3/4}}$. However, for states $\ket{\phi^{(0)}}$ with the same level as $\ket{\psi^{(0)}}$, which hence only differ in the zero-mode state\footnote{Note that states $\ket{\phi^{(0)}}$ with same level {\it and} number of zero-mode excitations as $\ket{\psi^{(0)}}$ do not couple in \eqref{eq:DoOm_2ndOrdPert}.}, one has $\calE^{(0)}_{\psi} - \calE^{(0)}_{\phi} \propto \l^{-1/4}$ such that the second order contribution to $\calE$ is indeed of order $\ord{\l^{-1/2}}$. By this, for general states $\ket{\psi^{(0)}}$ other modes giving rise to states with the same level do not decouple and will contribute to the energy at $\ord{\l^{-1/4}}$ via \eqref{eq:DoOm_2ndOrdPert}.

	However, acting with $\d H_{-3/8}$ on lowest excited non zero-mode states  
	 $\ket{\psi^{(0)}}=\alpha^i_{-1} \tilde{\alpha}^j_{-1} \ket{x_{0},p_{0}}$ inevitably changes the level and \eqref{eq:DoOm_2ndOrdPert} does not contribute to the energy at $\ord{\l^{-1/4}}$. Hence, we conclude that our previous results should be trusted for the Konishi state, $n=m=1$, only. {A similar observation has been made} in appendix A.1 of \cite{Passerini:2010xc}.

\bigskip

\section*{Acknowledgments}


\noindent
We thank Harald Dorn, Chrysostomos Kalousios, Thomas Klose, Joe Minahan, Jonas Pollok and Tristan McLoughlin for useful discussions.
G.J. thanks the Humboldt University Berlin and the Max-Planck Institute for Gravitational
Physics (Albert-Einstein-Institute) in Potsdam for warm hospitality.  M.H. thanks Nordita in Stockholm for kind hospitality.
The research leading to these results has received funding from 
the Volkswagen-Foundation, the International Max Planck Research School for Geometric Analysis, Gravitation and String Theory, and
the People Programme (Marie Curie Actions) of the European Union's Seventh Framework Programme FP7/2007-2013/ under REA Grant Agreement No 317089.
S.F. is supported by a DFG grant in the framework of the SFB 647 ÒRaum - Zeit - Materie. Analytische und Geometrische StrukturenÓ
and by the Science Foundation Ireland under Grant 09/RFP/PHY2142.

\newpage




\appendix

\section{ \texorpdfstring{$\text{SO}({N_q})$}{SO({N_q})} Gauging } \label{app:SONq}

Here we consider a ${N_q}$-dimensional harmonic oscillator and calculate mean values of SO$({N_q})$ scalar operators
for the states with vanishing angular momentum.

Let us introduce the  annihilation and creation operators
\be\label{eq:SONq_aadag}
a_\a=\frac{p_\a-iq_\a}{\sqrt 2}~, \qquad a^\dag_\a=\frac{p_\a+iq_\a}{\sqrt 2}~, \quad (\a=1,\dots ,{N_q})~,
\ee
which satisfy the standard commutation relations
$[a_\a, a_\b]=[a_\a^\dag, a_\b^\dag ]=0,$ $\,[a_\a, a_\b^\dag ]=\d_{\a\b}.$

Three operators constructed by quadratic scalar combinations in $\v a$ and $\v a^{\,\dag}$,
\be\label{eq:SONq_ONqScalars}
h=\frac{1}{2}(\v a^{\,\dag}\cdot\v a + \v a\cdot\v a^{\,\dag})~, \qquad A= \v a^{\,2}~, \qquad A^\dag = \v a^{\,\dag\,2}~,
\ee
form the ${\mathfrak su}(1,1)$ algebra
\be\label{eq:SONq_so21}
[h, A]=-2A~, \qquad [h, A^\dag]=2A^\dag~,\qquad [A,\,A^\dag]=4h~.
\ee
The corresponding Casimir relates to the total angular momentum operator ${\cal J}^2=\frac{1}{2}\,{\cal J}_{\a\b}{\cal J}_{\a\b}$ as
\be\label{eq:SONq_ScalarOpRels}
h^2-\frac{1}{2}(A\,A^\dag+A^\dag\,A)={{\cal J}^2}+\f{{N_q}^2}{4}-{N_q}~,
\ee
where the SO$({N_q})$ rotation generators ${\cal J}_{\a\b}=i(a^\dag _\a a_\b-a^\dag _\b a_\a)$ annihilate the states
\be\label{eq:SONq_ScalarStates}
|n\rangle \propto \, (A^\dag)^n |0\rangle~,
\ee
$|0\rangle$ being the standard vacuum.

Hence, $h |n\rangle=(2n+{N_q}/2) |n\rangle$, and
 by  \eqref{eq:SONq_ScalarOpRels} one finds
\be\label{eq:SONq_ExpVal-AAdag}
\langle n |(A\,A^\dag+A^\dag\,A)|n\rangle=8\,n^2+4\,{N_q}\,n+2{N_q}~.
\ee

With the help of \eqref{eq:SONq_so21}, from \eqref{eq:SONq_ScalarOpRels} we also obtain
\be\label{eq:SONq_ExciteAnnihilate}
A|n\rangle=\sqrt{2n(2n+{N_q}-2)}\,\,|n-1\rangle ~,  \quad A^\dag|n\rangle=
\sqrt{(2n+{N_q})(2n+2)}\,\,|n+1\rangle~,
\ee
which define the normalization of the states \eqref{eq:SONq_ScalarStates}.

By \eqref{eq:SONq_aadag}-\eqref{eq:SONq_ONqScalars} one has $\v p^{\,2}=h+\f{1}{2}(A+A^\dag)$, $\,\v q^{\,2}=h-\f{1}{2}(A+A^\dag)$, $D\equiv\frac{1}{2}(\v p\cdot\v q+\v q\cdot\v p)=\f{i}{2}(A-A^\dag),$ and using \eqref{eq:SONq_ScalarOpRels} once more,
one obtains the following mean values
\be\label{eq:SONq_MeanVals1}
\langle n |\v p^{\,2} |n\rangle=\langle n |\v q^{\,2} |n\rangle=2\,n+\f{N_q}{2}~,\qquad \quad \langle n |D^2 |n\rangle=
2\,n^2+{N_q}\,n+\f{N_q}{2}~,~~~~~~
\ee
\be\label{eq:SONq_MeanVals2}
\langle n | (\v q^{\,2})^2 |n\rangle
	=6\,n^2+3\,{N_q}\,n+\f{{N_q}^2}{4}+\f{N_q}{2}~,
\quad \langle n |\frac{1}{2}\left(\v p^{\,2}\v q^{\,2}+\v q^{\,2}\v p^{\,2}\right) |n\rangle
	=2\,n^2+{N_q}\,n+\f{{N_q}^2}{4}-\f{{N_q}}{2}~.
\ee
These equations show the dependence of mean values for different SO$({N_q})$ scalar operators
on the dimension of space ${N_q}$.


\bibliographystyle{nb}
\bibliography{bibSpires}


\end{document}